\documentstyle[12pt]{article}

\input{mssymb}

\font\ten=cmbx10 at 12pt

\renewcommand{\thefootnote}{\fnsymbol{footnote}}

\newcounter{saveeqn}

\newenvironment{goodlist}[1]%
{\begin{list}{}{\settowidth{\labelwidth}{#1}
  \setlength{\leftmargin}{\labelwidth}
  \addtolength{\leftmargin}{\labelsep}
  \setlength{\parsep}{1ex plus0.7ex minus0.7ex}
  \setlength{\itemsep}{0.8ex}
  }}{\end{list}}

\def\build#1_#2^#3{\mathrel{\mathop{\kern 0pt#1}\limits_{#2}^{#3}}}

\newcommand{\beq}[1]{ \begin{equation}\label{#1}}
\newcommand{\eeq}{\end{equation}}
\newcommand{\ba}{\begin{array}}
\newcommand{\ea}{\end{array}}
\newcommand{\beqa}{\begin{eqnarray}}
\newcommand{\eeqa}{\end{eqnarray}}

\begin{document}

\begin{titlepage}

\begin{center}

{\ten Centre de Physique Th\'eorique\footnote{Unit\'e Propre de
Recherche 7061} - CNRS - Luminy, Case 907}

{\ten F-13288 Marseille Cedex 9 - France }

\vspace{1 cm}

\par

{\large \bf IMPROVED VERSION OF THE EIKONAL MODEL
FOR ABSORBING SPHERICAL PARTICLES}

\vspace{1.0 cm}
\setcounter{footnote}{0}
\renewcommand{\thefootnote}{\arabic{footnote}}

{\bf Claude BOURRELY, Pierre CHIAPPETTA
 \\ and  \\
 Thierry LEMAIRE\footnote{Institut of Physics,
Federal University of Bahia, Salvador, Brazil}}
\vspace{1 cm}

{\bf Abstract}

\end{center}

We present a new expression of the scattering amplitude, valid
 for spherical absorbing objects,
which leads to an improved version of the eikonal method outside the
diffraction region. Limitations of this method are discussed and numerical
results
are presented and compared successfully with the Mie theory.

\vspace{1 cm}

\noindent Key-Words : Light scattering by spheres, eikonal approximation.

\bigskip

\noindent Number of figures : 7

\bigskip

\noindent April 1995

\noindent CPT-95/P.3186

\bigskip

\noindent anonymous ftp or gopher: cpt.univ-mrs.fr

\end{titlepage}

\section{Introduction}
The eikonal picture, or eikonal model, has been first developped by Perrin and
Chiappetta \cite{PerChia} to describe optical scattering phenomenons
by scatterers having an axis of symmetry parallel to the incident wave-vector.
This approximation is a variant of the High-Energy Approximation (HEA)
introduced by Glauber \cite{Glau} for scattering of particles by a
potential of infinite range. Even though this latter theory is valid only for
small scattering angle, the eikonal picture was proven to be valid
in the whole angular domain.
Numerical results \cite{PerChia}-\cite{PerLam1} for the sphere
shown rather good agreement with
the Mie theory, particularly when internal multiple reflexions are negligible.
This useful approximation provides a simple mathematical scheme and has been
used particularly in astronomy, to describe the light scattering by irregular
particles \cite{PerLam2}.

An other version of the eikonal method, called the generalized eikonal
approximation \cite{Chen}, including the second Born correction
 of the scattering
amplitude has been developped for the sphere, leading to good agreement with
the Mie theory. Recently, a vectorial version of the eikonal method valid
for objects of approximately spherical shape has been derived \cite{BCL1}.
The eikonal approximation has also been extended to the case of objects whose
axis of symmetry is not parallel to the direction of the incident wave-vector
\cite{BCL2}.

The eikonal picture applied to scattering by large spheres gives only
a qualitative description of the backward intensity.
In the present work, we show that simple considerations suggest a modification
of the eikonal model amplitude for large absorbing spheres, leading to a better
quantitative description of the scattering phenomena.
\par

\par
This paper is organized as follows : in section 2, we start by recalling the
eikonal model  and we give the scattering amplitude for an absorbing sphere.
In section 3, we consider this scattering amplitude in the forward and
backward limits. Comparisons with the diffraction and the optical
backscattering amplitudes suggest an empirical transformation of the scattering
amplitude which is tested against exact solutions in section 4.
The conclusion is given in section 5.

\section{The eikonal scattering amplitude for an absorbing sphere.}
\label{scampli}

The eikonal method leads to an approximate formulation of the solution
of the scalar Helmholtz equation
\beq{a1}
\Bigl \{ \Delta + m^2k^2 \Bigr \} \Psi (r) = 0,
\eeq
where $m$ is the refractive index of the scatterer, $\bf{k}$ is the wave-vector
of the incident plane wave, and $\Psi(r)$ is the electric field
perpendicular to the scattering plane \cite{PerLam1}.
For objects having an axis of symmetry parallel
to $\bf{k}$ the solution $\Psi (r)$ can be written for large $r$ :
\beq{a2}
\Psi (r) = e^{i\bf{k}\cdot \bf{r}} + f(\theta){e^{ikr} \over r}
 + O(\frac{1}{r^2}).
\eeq
The scattering amplitude is given \cite{PerChia}, in cylindrical
coordinates, by :
\beq{a3}
f(\theta) = - {k^2 \over 2} \int_{0}^{\infty} db b J_0(kb \sin \theta)
G(b) + O(1),
\eeq
where $G(b)$ is
\beq{a4}
G(b) = \int_{-\infty}^{\infty} dz U(b,z) e^{-iqz} \exp{ \Bigl \{
i {k \over 2} \int_{-\infty}^{z(b)} dz' U(b,z') \Bigr \} },
\eeq
$z(b)$ is the boundary of the scatterer for a given value of the
impact parameter $b$, and $U(b, z) = 1 - m^2$, $q = 2k \sin^2{
{\theta \over 2}}$. The incident wave-vector $\bf{k}$ has been taken
 parallel to the $z$ axis (see Fig. 1). The function $G(b)$ can be
computed exactly when the refractive index is a constant.
For a convex particle whose center of symmetry is located at the origin :
\beq{a5}
G(b) = { 2 \over k}\gamma (\theta) \Bigl \{ e^{-iq z(b) + i\chi(b)}
- e^{iq z(b)} \Bigr \},
\eeq
with $\gamma(\theta) = \displaystyle{{ kU \over {2q - kU}}}$ and
the eikonal function $\chi(b)$ reads :
\beq{a6}
\chi(b) = {k \over 2} \int_{-\infty}^{z(b)} dz' U(b, z').
\eeq
For a sphere of radius $a$, $\chi(b) = k U z(b)$. Then the scattering
amplitude has the simple expression :
\beq{a7}
f(\theta) \approx -i k \gamma(\theta) \int_0^a db b J_0(k b \sin{\theta})
\Bigl \{  e^{-iq z(b) + i\chi(b)}- e^{iq z(b)} \Bigr \},
\eeq
with $z(b) = \sqrt{a^2 -b^2}$.

\noindent When the refractive index has a non negligible imaginary part,
{\it i.e.} when the condition $|Im(m)| k a \gtrsim 1$ is satisfied, the real
part of $i \chi(b)$ takes a high negative value which leads to the approximate
expression of $f(\theta)$, namely :
\beq{a8}
f(\theta) \approx i k \gamma(\theta) \int_0^a db b J_0(k b \sin{\theta})
e^{iq z(b)}.
\eeq

\section{Modification of the eikonal model}
\label{modif}

Assuming that the sphere is sufficiently absorbing means that physically,
the diffractive and the first order reflective parts of the
 scattering amplitude give a reasonable description of the scattering
phenomenon. The present expression (\ref{a7}) of the scattering
amplitude leads only to qualitative results in the backward direction
 and needs to be modified. Let us first consider the
behaviour of $f(\theta)$ in (\ref{a8}) when $\theta$ is small
(~{\it i.e.} when the diffractive part dominates). Since,
$e^{iq z(b)} \approx 1$, we obtain~:
\beq{a9}
 f(\theta) \approx {i a \over \sin{\theta}} J_1(ka \sin{\theta}),
\eeq
showing that we recover
the exact expression of the diffraction amplitude for a sphere.
 On the other hand, in the neighbourhood of the backscattering angle
( {\it i.e.} $|\pi - \theta | << 1$), the use of the method of stationary
phase leads to :
\beq{a10a}
f(\theta) \approx {a \over 2 \sin^2{{\theta \over 2}}}
\exp{(2 i k a \sin^2{{\theta \over 2}})} \gamma(\theta).
\eeq
Since $\sin^2{{\theta \over 2}}$ is close to 1, we approximate (\ref{a10a})
as :
\beq{a10}
f(\theta) \approx {a \over 2} \exp{(2 i k a \sin{{\theta \over 2}})}
\gamma(\theta).
\eeq
Whereas the geometrical optics gives :
\beq{a11}
f_{Opt} \approx {a \over 2} \exp{(2 i k a \sin{{\theta \over 2}})}
 r_\perp(\theta),
\eeq
where $r_\perp(\theta)$ \cite{VanH} is the Fresnel reflection coefficient
for the perpendicular component of the electric field, which was proven
in \cite{PerLam1} and \cite{BCL1} to be identical to the eikonal amplitude.
We observe that the expression (\ref{a10}) differs from (\ref{a11}) by a
factor $\displaystyle{\gamma(\theta) \over r_\perp(\theta)}$
which has a modulus
different from 1 for refractive index near 1, leading for large angle
to a systematic error in the evaluation of the scattering intensity.
However, since $r_\perp(0) = \gamma(0)$,
a simple modification of the eikonal model which takes into account
the forward and the backward
asymptotic behaviors (\ref{a9}) and (\ref{a11}), would consist in
replacing $\gamma(\theta)$ by $r_\perp(\theta)$, leading to~:
\beq{a12}
f_{mod}(\theta) \approx i k r_\perp(\theta) \int_0^a db b
J_0(k b \sin{\theta}) e^{i q z(b)}.
\eeq
This transformation has the advantage of being able to reproduce the
two main components of the scattering amplitude valid respectively
in the forward and the backward scattering directions.
The comparison between the modified scattering intensity and the Mie theory,
will be given in the next section.

\section{Numerical results}
\label{numer}

In this section, we will compute the scattering intensity
$i(\theta) = k^2 |f(\theta)|^2$ for homogeneous, large spheres ($ka \geq 20$),
from the original eikonal model (\ref{a7}) and  the modified one (\ref{a12}).
The model is in fact valid for $k a \geq 10$, and according to a numerical
point of view, the new version is strictly equivalent to the original one
when the expression (\ref{a8}) is used.
The figures (2-6) show the results
obtained when the refractive index has an imaginary part sufficiently
large (according to the relation written above).
We observe that the modified eikonal formula (dashed curves) is in much
better agreement with the Mie theory (solid curves) for scattering
angles greater than $40^o$, as long as $|Im(m)| k a \gtrsim 1$ and
 $k a >> 1$. Indeed, in this angular region, since the reflection
 becomes the main component of the scattering amplitude, the intensity
predicted by the original eikonal model (dotted curves) lies systematically
above the Mie theory. We observe that no oscillations appear for angles
more or less greater than $40^o$
because the internal multiple reflections in the sphere are not included
in the new formulation, leading to a smooth description of the
scattered intensity clearly seen in Fig. 6. In Fig. 7, we have considered
the case of an ellipsoid with $ka = 50$ and $kb = 75$ in the $z$ direction.
Although no comparison can be made with respect to the Mie theory, there is
again a difference between the modified eikonal model (dashed curve) and
the original one (dotted curve) for scattering angles above $40^0$.
A comparison with a sphere of the same $ka$ and index (Fig. 1) shows that the
scattered intensity for the ellipsoid is lower than the former in the
backward direction, an effect partially due to more absorption.

\section{Conclusion}
\label{conclu}

We have derived a modified version of the eikonal model valid for large
absorbing spheres ($|Im(m)| k a \gtrsim 1, k a >> 1$),
by imposing that the asymptotic expressions
of the eikonal model amplitude in the forward and backward limits
reproduce the diffractive scattering amplitude
and the backscattered amplitude obtained from geometrical optics.
We have shown that one has to replace the coefficient
$\gamma(\theta)$ in the formula (\ref{a8}) by the Fresnel reflection
coefficient $r_\perp(\theta)$.
Numerical results show a great improvement of the eikonal
formulation, especially for scattering angles greater than $40^o$ and
a much closer agreement with the Mie theory.

\vskip 1cm
\noindent T. Lemaire thanks the Brazilian agency CNPq for his financial
support (project 300792/93-0).

\newpage

\section*{Figure Captions}

\begin{goodlist}{Fig.6}

\item[Fig.1]Geometry and notations.

\item[Fig.2]Scattering intensity for a sphere of size parameter
$ka = 50$ and refractive index $m = 1.33 + i0.05$. Mie theory (solid),
modified eikonal model (dashed), original eikonal  model (dotted).

\item[Fig.3]Scattering intensity for a sphere of size parameter
$ka = 20$ and refractive index $m = 1.5 + i0.1$. Mie theory (solid),
modified eikonal model (dashed), original eikonal model (dotted).

\item[Fig.4]Scattering intensity for a sphere of size parameter
$ka = 60$ and refractive index $m = 2.5 + i0.05$. Mie theory (solid),
modified eikonal model (dashed), original eikonal model (dotted).

\item[Fig.5]Scattering intensity for a sphere of size parameter
$ka = 20$ and refractive index $m = 5.0 + i0.1$. Mie theory (solid),
modified eikonal model (dashed), original eikonal model (dotted).

\item[Fig.6]Scattering intensity for a sphere of size parameter
$ka = 100$ and refractive index $m = 2.5 + i0.01$. Mie theory (solid),
modified eikonal model (dashed), original eikonal model (dotted).

\item[Fig.7]Scattering intensity for an ellipsoid of size parameter
$ka = 50$, $kb = 75$ and refractive index $m = 1.33 + i0.05$.
Modified eikonal model (dashed), original eikonal model (dotted).

\end{goodlist}


\begin{thebibliography}{99}

\bibitem{PerChia}PERRIN J.M. and CHIAPPETTA P., 1985, {\it Opt. Acta},
{\bf 32}, 907.

\bibitem{Glau}GLAUBER R.J., 1958, {\it Lectures in Theoretical Physics},
{\bf vol. 1}, 315, ed. Brittin W.L. and Dunham L.G.
(New York: Interscience).

\bibitem{PerLam1}PERRIN J.M. and LAMY P.L., 1986, {\it Opt. Acta},
{\bf 33}, 1001.

\bibitem{PerLam2}PERRIN J.M. and LAMY P.L., 1983, {\it Opt. Acta},
{\bf 30}, 1223

BOURRELY C., CHIAPPETTA P. and TORRESANI B., 1986 {\it J. Opt. Soc. Am.},
{\bf A3}, 250

BOURRELY C., CHIAPPETTA P. and TORRESANI B., 1986 {\it Opt. Comm.},
{\bf 58}, 365.

\bibitem{Chen}CHEN T.W., 1989, {\it Appl. Opt.}, {\bf 28}, 4096

CHEN T.W. and SMITH W.S., 1992, {\it Appl. Opt.}, {\bf 31}, 6558.

\bibitem{BCL1}BOURRELY C., CHIAPPETTA P. and LEMAIRE T., 1991,
 {\it J. Mod. Opt.}, {\bf 38}, 305.

\bibitem{BCL2}BOURRELY C., CHIAPPETTA P. and LEMAIRE T., 1989,
 {\it Opt. Comm.}, {\bf 70}, 173.

\bibitem{VanH}VAN de HULST H.C., 1981, {\it Light Scattering by Small
 Particles} (New York: Dover Publications, Inc.).

\end{thebibliography}
\end{document}